\documentclass[10pt,journal,compsoc]{IEEEtran}
\usepackage{graphicx}
\usepackage{booktabs}
\usepackage{amsmath}
\usepackage{multirow}

\usepackage{color,soul}

\usepackage[T1]{fontenc}

\usepackage{cite}

\ifCLASSINFOpdf
\else
\fi

\begin{document}

\title{Tactile Roughness Perception of \hspace{\textwidth}Virtual Gratings by Electrovibration}

\author{Aykut~\.{I}\c{s}leyen,
        Yasemin~Vardar, \IEEEmembership{Member,~IEEE},
        and~Cagatay~Basdogan, \IEEEmembership{Senior Member,~IEEE}% <-this % stops a space
\IEEEcompsocitemizethanks{\IEEEcompsocthanksitem A. \.{I}\c{s}leyen and C. Basdogan are with the Department of Mechanical Engineering,
Ko\c{c} University, Istanbul 34450, Turkey.\protect\\
% note need leading \protect in front of \\ to get a newline within \thanks as
% \\ is fragile and will error, could use \hfil\break instead.
E-mail: {aisleyen16,cbasdogan}@ku.edu.tr.
\IEEEcompsocthanksitem Y. Vardar is with the Haptic Intelligence Department, Max Planck Institute for Intelligent Systems, Stuttgart 70569, Germany.\protect\\
E-mail: yvardar@is.mpg.de.
}% <-this % stops an unwanted space
}

% The paper headers
\markboth{IEEE TRANSACTIONS ON HAPTICS, VOL. NN, NO. NN, MONTH YEAR}%
{}%

\IEEEtitleabstractindextext{%
\begin{abstract}

Realistic display of tactile textures on touch screens is a big step forward for haptic technology to reach a wide range of consumers utilizing electronic devices on a daily basis. Since the texture topography cannot be rendered explicitly by electrovibration on touch screens, it is important to understand how we perceive the virtual textures displayed by friction modulation via electrovibration. We investigated the roughness perception of real gratings made of plexiglass and virtual gratings displayed by electrovibration through a touch screen for comparison. In particular, we conducted two psychophysical experiments with 10 participants to investigate the effect of spatial period and the normal force applied by finger on roughness perception of real and virtual gratings in macro size. We also recorded the contact forces acting on the participants' finger during the experiments. The results showed that the roughness perception of real and virtual gratings are different. We argue that this difference can be explained by the amount of fingerpad penetration into the gratings. For real gratings, penetration increased tangential forces acting on the finger, whereas for virtual ones where skin penetration is absent, tangential forces decreased with spatial period. Supporting our claim, we also found that increasing normal force increases the perceived roughness of real gratings while it causes an opposite effect for the virtual gratings. These results are consistent with the tangential force profiles recorded for both real and virtual gratings. In particular, the rate of change in tangential force ($dF_t/dt$) as a function of spatial period and normal force followed trends similar to those obtained for the roughness estimates of real and virtual gratings, suggesting that it is a better indicator of the perceived roughness than the tangential force magnitude.

\end{abstract}

}

\maketitle
\IEEEdisplaynontitleabstractindextext
\IEEEpeerreviewmaketitle
\IEEEraisesectionheading{\section{Introduction}}

\IEEEPARstart{T}{ouch} screens have been used in a wide range of portable devices nowadays, but our interactions with these devices mainly involve visual and auditory sensory channels. While a commercial touch screen today can easily detect finger position and hand gestures, it provides limited tactile feedback. However, tactile feedback can be used as an additional sensory channel to convey information and also reduce the perceptual and cognitive load on the user. Currently, friction modulation is the most promising approach to display tactile feedback through a touch screen. In this regard, there are two promising techniques: ultrasonic and electrostatic actuation. In the case of ultrasonic actuation \cite{61,65,79,60,64,95,96,97}, the surface is vibrated at an ultrasonic resonance frequency. As a result, a squeezed thin film of air between finger and the surface is formed. This layer breaks the direct contact of finger with the surface and hence leads to a reduction in friction. On the other hand, electrostatic actuation  \cite{2,1,4,5,12} increases the friction between finger and surface by electroadhesion. When an alternating voltage is applied to the conductive layer of a capacitive touch screen, an attractive electrostatic force is generated in the normal direction between the finger and the surface. By controlling the amplitude, frequency, and waveform of the input voltage, the frictional force between the sliding finger and the touch screen can be modulated. This technology has a great potential especially in mobile applications including online shopping, games, education, data visualization, and development of aids for blind and visually impaired. In this context, one important aim is to render realistic virtual textures on touch screens. Texture information on touch screens would improve the user experience in daily activities. For example, feeling the simulated texture of a jean before purchasing it from Internet would certainly be more motivating for online shoppers. However, our knowledge on tactile perception of virtual textures displayed by friction modulation is quite limited though tactile perception of real textures has been already investigated extensively in the literature. Based on the multi-dimensional-scaling (MDS) studies conducted by Hollins et. al. \cite{29,45}, there are three independent perceptual dimensions in texture perception: roughness, hardness, and warmness. Among the three dimensions, roughness is arguably the most important dimension in tactile perception of textures.

To investigate the roughness perception of real textures, several types of stimuli have been used; raised dots with controlled height and density \cite{90,36}, dithered cylindrical raised elements \cite{18,26,19}, and metal plates with linear gratings \cite{20,23,30}.  These studies have shown that size of the tactile elements (i.e gratings, dots, cones) and the spacing between them are critical parameters in roughness perception. Moreover, Hollins et al. \cite{19} and Klatzky and Lederman \cite{52} found that the underlying mechanism behind roughness perception is different for micro-textures (textures having inter-element spacing approximately smaller than 0.2 mm) and macro-textures (textures having inter-element spacing approximately larger than 0.2 mm). At macro-textural scale, Lederman and colleagues \cite{18,20,23,30} observed that groove width (GW) has a greater effect on perceived roughness than ridge width (RW). This observation has been supported by other studies later \cite{90,36,46} reporting that the perceived roughness increases with the groove width (and hence with the spatial period) until it saturates.

In contrast to the extensive literature on real textures \cite{29,45,90,36,18,26,19,52,20,23,30,46,98}, the number of studies investigating the roughness perception of virtual textures rendered on a touch surface by electrovibration is limited. The studies in this area have mainly focused on the estimation of perceptual thresholds for periodic stimuli so far, but not their roughness perception. Bau et al. measured the sensory thresholds of electrovibration using sinusoidal inputs applied at different frequencies \cite{2}. They showed that the change in threshold voltage as a function of frequency followed a U-shaped curve similar to the one observed in vibrotactile studies. Later, Wijekoon et al. \cite{12}, followed the work of \cite{3}, and investigated the perceived intensity of friction generated by electrovibration. Their experimental results showed that the perceived intensity was logarithmically proportional to the amplitude of the applied voltage signal.

Additionally, there are also studies that investigate the underlying perceptual mechanism of virtual textures. Vardar et al. \cite{4,51} studied the effect of input voltage waveform on our tactile perception of electrovibration. Through psychophysical experiments with 8 subjects, they showed that humans were more sensitive to tactile stimuli generated by square wave voltage than sinusoidal one at frequencies below $60$ Hz. They showed that Pacinian channel was the primary psychophysical channel in the detection of the electrovibration stimuli, which is most effective to tactile stimuli at frequencies around  $250$ Hz. Hence, the stronger tactile sensation caused by a low-frequency square wave was due to its high-frequency components stimulating the Pacinian channel.

There are only a few studies in the literature on roughness perception of virtual gratings rendered by electrovibration. Ilkhani et. al \cite{67} conducted multidimensional scaling analysis (MDS) on data-driven textures taken from Penn Haptic Texture Toolkit \cite{73} and concluded that roughness is one of the main dimensions in tactile perception of virtual textures. Vardar et al. \cite{53} investigated the roughness perception of four waveforms; sine, square, triangular and saw-toothed waves with spatial period varying from $0.6$ to $8$ mm. The width of periodic high friction regimes (analogous to ridge width) was taken as $0.5$ mm, while the width of the low friction regimes (analogous to groove width) was varied. The finger velocity was controlled indirectly by displaying a visual cursor moving at $50$ mm/s. The results showed that square waveform was perceived as the roughest, while there was no significant difference between the other three waveforms. Vardar et al. \cite{89} also investigated the interference of multiple tactile stimuli under electrovibration. This interference is called tactile masking and can cause deficits in perception. They showed that sharpness perception of virtual edges displayed on touch screens depends on the "haptic contrast" between background and foreground tactile stimuli, which varies as a function of masking amplitude and activation levels of frequency-dependent psychophysical channels. This outcome suggests that tactile perception of virtual gratings can be altered by masking since they are constructed by a series of rising and falling virtual edges.

As it is obvious from the above paragraph, the number of studies on roughness perception of virtual textures rendered on a touch surface by electrovibration is only a few and the underlying perceptual mechanisms have not been established yet. In this study, we investigate how the perceived roughness of real and virtual square gratings change as a function of spatial period and normal force applied by finger to the touch screen. Earlier studies on real gratings \cite{18,19,22,24,21,25} mostly investigated the roughness perception but not paid sufficient attention to the contact interactions. However, it was shown in \cite{20} and \cite{23} that perceived roughness increases with increasing normal force applied by finger. Also, Taylor and Lederman \cite{28} showed that skin penetration into the inter-element spacing might predict the perceived roughness as a function of groove width and normal force. Since friction modulation displays cannot explicitly render surface topography in the normal direction, it is expected that perception of virtual gratings do not perfectly match their real counterparts. Our results also suggest that perception of real and virtual textures is mediated by different mechanisms.

\section{Virtual Texture Rendering}

The virtual textures should be rendered as realistically as possible to be able to investigate the perceptual differences between them and their real counterparts systematically. However, the best method for rendering realistic virtual textures on touchscreens has yet to be developed. In this section, we explain our virtual texture rendering method. In order to render virtual square gratings that mimic the real ones, we first investigated the contact interactions between human finger and real square gratings. For this purpose, we recorded the contact forces and analysed them in both time and frequency domain.

\begin{figure}[!b]
\centering
\includegraphics[width=3.5 in]{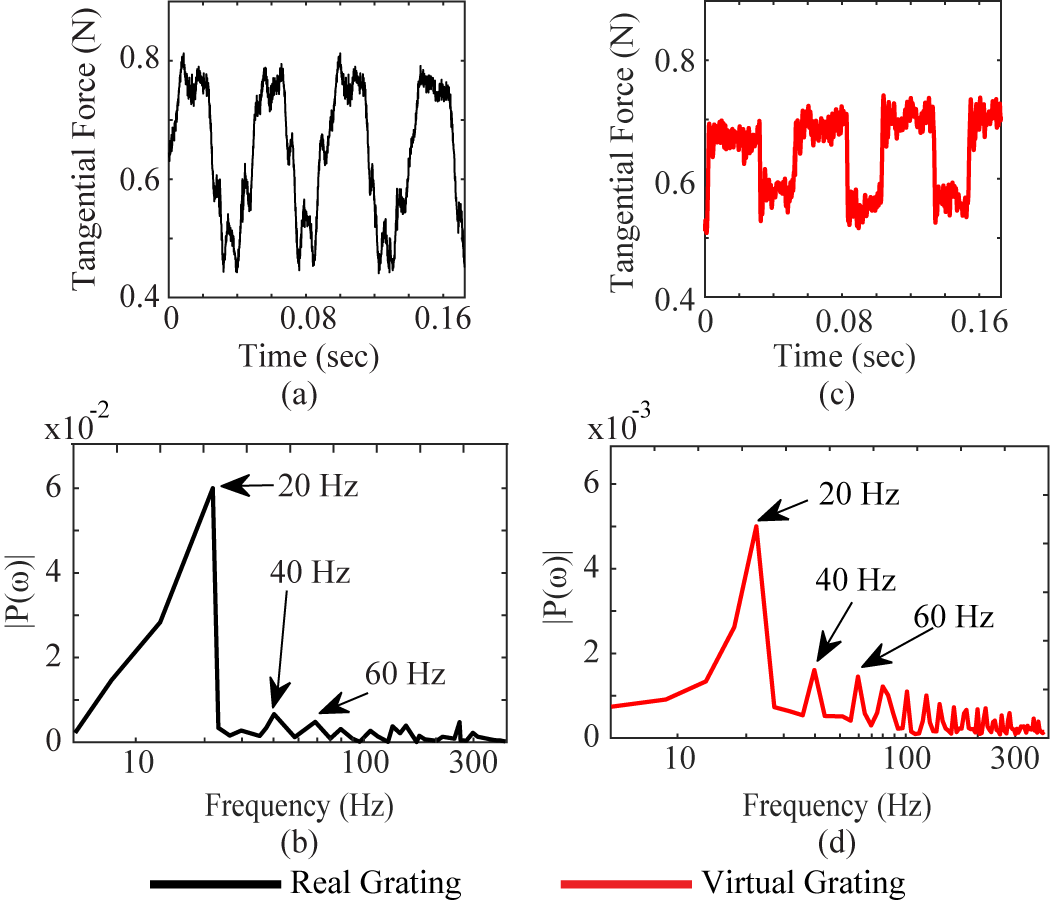}%
\caption{(a) Friction force signal acquired from a real grating (spatial period $=$ $2.5$ mm) under the constant normal force of $0.75$ N and the targeted exploration speed of $50$ mm/s. The period of the signal is approximately $0.05$ sec, which corresponds to the temporal frequency of $20$ Hz ($50/2.5$). (b) Power spectrum of the tangential force signal shown in (a); peaks appear at the integer multiples of the temporal frequency. (c) Tangential force signal acquired from the corresponding virtual grating displays high and low friction regimes. (d) Power spectrum of the tangential force signal shown in (c).}
\label{fig:realforce}
\end{figure}

We manufactured square gratings from plexiglass using a laser cutter in different groove widths. We fixed the ridge width of the gratings as $1$ mm and varied the groove width from $1.5$ mm to $7.5$ mm (corresponds to varying spatial period from $2.5$ mm to $8.5$ mm) to produce $6$ different gratings, similar to the ones utilized in the earlier texture studies (see Table \ref{table:stimulus2}). To analyze the frequency spectrum of contact forces, we selected one of the gratings with a spatial period of $2.5$ mm and recorded the frictional forces acting on the finger of one participant (i.e. the experimenter) while he slides his finger on the grating with a velocity of $50$ mm/s under a constant normal force of $0.75$ N.  As shown in Fig. \ref{fig:realforce}a, the period of the tangential force signal was $0.05$ sec, corresponding to a temporal frequency of $20$ Hz ($50/2.5$). Hence, the finger spends 0.02 and 0.03 secs on each ridge and groove respectively, leading to a duty cycle of $0.4$ ($0.02/0.05$). The power spectrum of the force signal (Fig. \ref{fig:realforce}b) revealed a series of peaks with decreasing magnitude at frequencies that are integer multiples of the temporal frequency. This spectrum resembles to the power spectrum of a pulse train signal. In order to generate virtual gratings having the similar frequency spectrum of the real gratings, we modulated a low frequency pulse train voltage with a high frequency carrier voltage signal as suggested in \cite{56,82,93}. As discussed by Shultz et. al \cite{82}, impedance of the gap between fingerpad and touch screen causes a volatile transition of force dynamics in the frequency range of $20 - 200$ Hz. They argued that amplitude modulation with a high frequency carrier voltage signal avoids this transition regime and hence, the modulated voltage signal results in a tangential force signal with a rectified DC component coming from the envelope signal and an AC component coming from the carrier signal (Fig. \ref{fig:realforce}c). If the frequency of carrier signal is selected as higher than the human vibrotactile threshold level of $1$ kHz \cite{57,58}, then the AC component of the resulting tangential force is not perceived by the user.

\begin{figure}[!t]
\centering
\includegraphics[width=3.5 in]{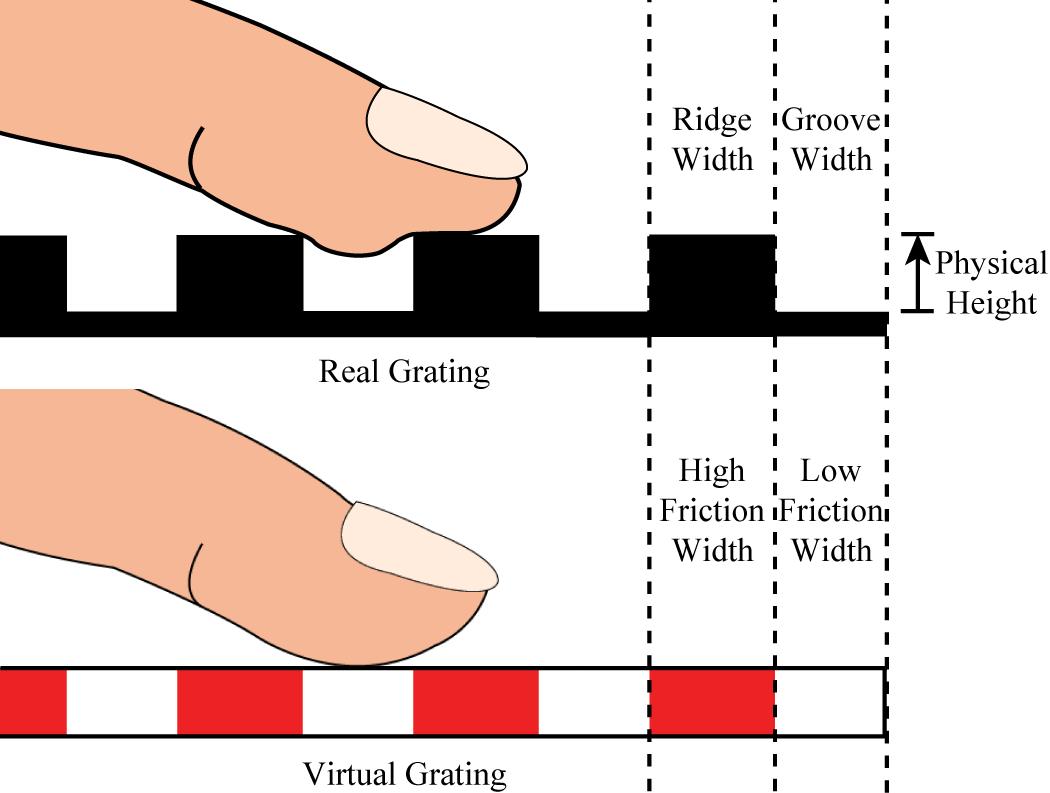}%
\caption{Grooves and ridges of a real grating are rendered as high and low friction widths (regimes) in the corresponding virtual grating, respectively. Finger penetrates into the real grating but not the virtual grating since grating height cannot be rendered explicitly by electrovibration. In our approach, the virtual gratings are designed in temporal domain and not in spatial domain, hence the finger cannot be partly on a high friction zone and partly on a low friction zone.}
\label{fig:periyodik-real-virtual}
\end{figure}

The signal modulation technique discussed above, in fact, creates periodic widths of high and low friction regimes (zones) on the surface of touch screen (Fig. \ref{fig:periyodik-real-virtual}). For example, if we design a virtual grating using the envelope frequency of $20$ Hz, duty cycle of $0.4$ (high friction width/spatial period), and carrier frequency of $3$ kHz, then the frequency spectrum of the resulting tangential force signal (Fig. \ref{fig:realforce}d) resembles to the one observed for the real grating (Fig. \ref{fig:realforce}b). However, we should note that the resulting high and low friction zones in our design depend on the selected exploration speed. If the exploration speed is changed, the temporal signals should be redesigned accordingly.

\section{Psychophysical Experiments}

We conducted psychophysical experiments on roughness perception of real and virtual gratings. In particular, we investigated the effect of spatial period and normal force on perceived roughness of real and virtual gratings. We initially aimed to conduct the experiments under $3$ different normal forces ($0.25$, $0.75$, $1.75$ N) and $6$ different spatial periods ($2.5$, $3$, $3.5$, $4.5$, $6.5$, $8.5$ mm), displayed multiple times in random order. However, during our preliminary studies, we observed that the tactile sensitivity of the participants was reduced due to finger wear when the number of trials was high. Hence, we simplified our experimental design and divided the experiments into two sets, executed in multiple sessions in different days to prevent finger wear. These two sets of experiments were performed for both real and virtual gratings for comparison. In the first set (Exp. 1), we conducted roughness estimation experiments for $6$ different spatial periods ($2.5$, $3$, $3.5$, $4.5$, $6.5$, $8.5$ mm) under the normal force of $0.75$ N. In the second set (Exp. 2), we conducted roughness estimation experiments for $3$ different normal forces ($0.25$, $0.75$, $1.75$ N) under $2$ spatial periods of $2.5$ mm and $8.5$ mm. 

\subsection{Experimental Setup}
\begin{figure}[!t]
\centering
\includegraphics[width=3.5 in]{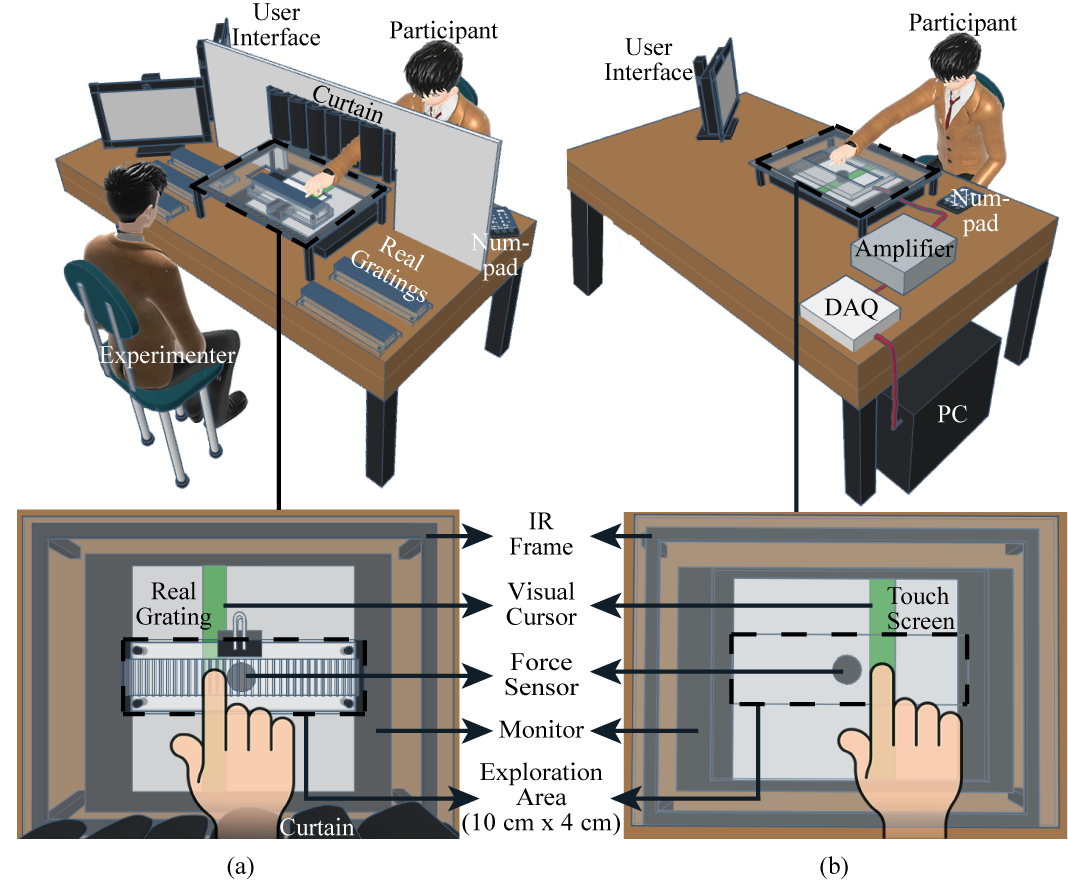}%
\caption{The experimental setup for investigating tactile 
roughness perception of real (a) and virtual (b) gratings.}
\label{fig:exp_set}
\end{figure}

The experimental setups for investigating the roughness perception of real and virtual gratings were slightly different (Fig. \ref{fig:exp_set}). For both setups, a compact monitor displaying a visual cursor moving with a speed of $50$ mm/s was placed below the real grating surface and the virtual one displayed through the touch screen to adjust the exploration speed of the participants. A force sensor (Nano17, ATI Inc.) was also placed under the grating surface to acquire normal and tangential forces in each trial. The force sensor had a sampling rate of $10$ kHz. In addition, an IR frame with a positional resolution of $1$ mm and a sampling rate of $85$ Hz was placed above the grating surface to monitor the exploration speed of the participants.

To display virtual gratings, a surface capacitive touch screen (SCT3250, 3M Inc.) with dimensions of $20$ x $15$ cm was used (see Fig. \ref{fig:exp_set}b). Voltage signal applied to the conductive layer of touch screen was generated by a DAQ card (USB-6251, National Instruments Inc.) working at a sampling rate of $10$ kHz. The signal was boosted by an amplifier (E-413, PI Inc.) before transmitted to the touch screen. As mentioned earlier, the real gratings were manufactured from plexiglass using a laser cutter. Each real grating surface had a length of $100$ mm and a width of $30$ mm.

\subsection{Participants}
Both experiments (Exp. 1, Exp. 2) were conducted with $2$ different groups of $10$ participants. The average ages of the participants in Exp. 1 and Exp. 2 were $24.9$ $\pm$ $1.3$ and $24.3$ $\pm$ $1.2$, respectively. Both groups were made of $5$ male and $5$ female participants. All participants were senior undergraduate or graduate university students and right-handed. They washed their hands with soap and rinsed with water before each session of the experiment. Moreover, their index finger and the touch screen were cleaned using ethanol before each session. Participants read and signed the consent form approved by Ethical Committee for Human Participants of Ko\c{c} University.

\subsection{Experimental Procedure}
In both experiments, the same experimental protocol was followed. The participants were instructed to sit on a chair and move their index fingers on the grating in tangential direction back and forth only once while synchronizing their finger movements with the movement of a visual cursor displayed by a monitor (see Fig. \ref{fig:exp_set}). The participants were allowed to replay each stimulus only once by pressing the '$0$' button on a numpad. To prevent any noise affecting their tactile perception, they were asked to wear noise cancellation headphones. Before the experiment, the participants were given instructions about the experiment and presented with a training session displaying all stimuli of that session once in random order. In both experiments, it took about $30$ minutes for each participant to complete one session including training. If the magnitude of mean normal force applied by the participant was $\pm30\%$ off the desired value, the participants were prompted to repeat the trial. In the case of real gratings, the participants sat on a chair at a table and extended their dominant arm under a curtain that prevented them from seeing the grating surface and the experimenter seated on the opposite site (see Fig. \ref{fig:exp_set}a). The experimenter manually changed the real grating surfaces during the experiments and it took around $3$ seconds to make the change. On the other hand, the virtual gratings were displayed automatically by the computer after each trial with no external intervention by the experimenter (Fig. \ref{fig:exp_set}b). After each trial, the participants entered their ratings of the stimulus using a small numpad. In both cases (real and virtual), participants were allowed to enter any positive number as their magnitude estimation of tactile roughness. They could see their responses on the user interface and could change it until they hit the 'return' button. After hitting the 'return' button on numpad, a new grating was displayed to the participants for exploration.

\subsection{Experiment 1}
In Exp. 1, we investigated the effect of spatial period on roughness perception of real and virtual gratings separately for the normal force of $0.75$ N.

\subsubsection*{Stimuli}
We selected the spatial periods of the real and virtual gratings in reference to the earlier studies on real gratings (see Table \ref{table:stimulus2}). The voltage signal for virtual gratings was generated using the amplitude modulation technique discussed in Section 2. The frequency of the carrier signal was fixed at $3$ kHz, but the frequency of the envelope signal and the duty cycle were set according to the desired spatial period (see Table \ref{table:stimulus2}). The number of real and virtual gratings displayed separately to participants was $108$ ($6$ spatial periods x $6$ repetitions x $3$ sessions). Hence, each session of the experiments for real and virtual gratings consisted of $36$ stimuli.

\begin{table}[!h]
\caption{Experimental parameters and their corresponding values\hspace{\textwidth}used in Exp. 1 (GW: Groove Width, RW: Ridge Width.)}
\centering
\renewcommand{\arraystretch}{1.5}
\label{table:stimulus2}
\small
\resizebox{\columnwidth}{!}{
\renewcommand{\arraystretch}{1.3}

\begin{tabular}{ccclcc}
\cline{1-3} \cline{5-6}
\multicolumn{3}{|c|}{\begin{tabular}[c]{@{}c@{}}Real Gratings\end{tabular}}                                                                                                                                               & \multicolumn{1}{l|}{} & \multicolumn{2}{c|}{\begin{tabular}[c]{@{}c@{}}Virtual Gratings\end{tabular}}                                                             \\ \cline{1-3} \cline{5-6} 
\multicolumn{6}{l}{}                                                                                                                                                                                                                                                                                                                                                                                                                 \\ \\[-0.33in] \cline{1-3} \cline{5-6} 
\multicolumn{1}{|c|}{\begin{tabular}[c]{@{}c@{}}Spatial Period\\ (mm)\end{tabular}} & \multicolumn{1}{c|}{\begin{tabular}[c]{@{}c@{}}GW\\ (mm)\end{tabular}} & \multicolumn{1}{c|}{\begin{tabular}[c]{@{}c@{}}RW\\ (mm)\end{tabular}} & \multicolumn{1}{l|}{} & \multicolumn{1}{c|}{\begin{tabular}[c]{@{}c@{}}Duty\\ Cycle\end{tabular}} & \multicolumn{1}{c|}{\begin{tabular}[c]{@{}c@{}}Envelope\\ Frequency\\ (Hz)\end{tabular}} \\ \cline{1-3} \cline{5-6} 
\multicolumn{1}{|c|}{2.5}                                                           & \multicolumn{1}{c|}{1.5}                                               & \multicolumn{1}{c|}{\multirow{6}{*}{1}}                                & \multicolumn{1}{l|}{} & \multicolumn{1}{c|}{0.4}                                                  & \multicolumn{1}{c|}{20}                                                                  \\ \cline{1-2} \cline{5-6} 
\multicolumn{1}{|c|}{3}                                                             & \multicolumn{1}{c|}{2}                                                 & \multicolumn{1}{c|}{}                                                  & \multicolumn{1}{l|}{} & \multicolumn{1}{c|}{0.33}                                                 & \multicolumn{1}{c|}{16.67}                                                               \\ \cline{1-2} \cline{5-6} 
\multicolumn{1}{|c|}{3.5}                                                           & \multicolumn{1}{c|}{2.5}                                               & \multicolumn{1}{c|}{}                                                  & \multicolumn{1}{l|}{} & \multicolumn{1}{c|}{0.29}                                                 & \multicolumn{1}{c|}{14.29}                                                               \\ \cline{1-2} \cline{5-6} 
\multicolumn{1}{|c|}{4.5}                                                           & \multicolumn{1}{c|}{3.5}                                               & \multicolumn{1}{c|}{}                                                  & \multicolumn{1}{l|}{} & \multicolumn{1}{c|}{0.22}                                                 & \multicolumn{1}{c|}{11}                                                                  \\ \cline{1-2} \cline{5-6} 
\multicolumn{1}{|c|}{6.5}                                                           & \multicolumn{1}{c|}{5.5}                                               & \multicolumn{1}{c|}{}                                                  & \multicolumn{1}{l|}{} & \multicolumn{1}{c|}{0.15}                                                 & \multicolumn{1}{c|}{7.67}                                                                \\ \cline{1-2} \cline{5-6} 
\multicolumn{1}{|c|}{8.5}                                                           & \multicolumn{1}{c|}{7.5}                                               & \multicolumn{1}{c|}{}                                                  & \multicolumn{1}{l|}{} & \multicolumn{1}{c|}{0.12}                                                 & \multicolumn{1}{c|}{5.88}                                                                \\ \cline{1-3} \cline{5-6} 
\end{tabular}
}
\end{table}	

\begin{figure*}[!ht]
\centering
\includegraphics[width=7 in]{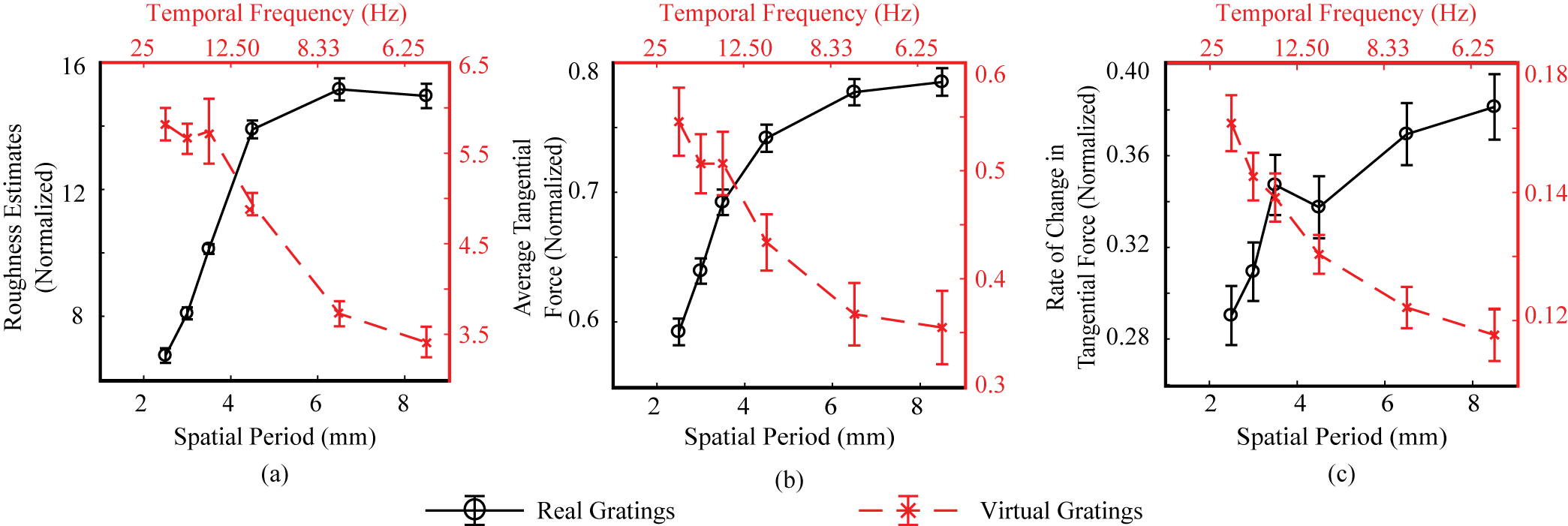}%
\caption{Results of Exp. 1; normalized roughness estimates (means and standard mean errors), average tangential force ($F_t$), and rms of rate of change in tangential force ($dF_t/dt$) for real (solid - black) and virtual (dashed - red) gratings under the normal force of $0.75$ N and exploration speed of $50$ mm/s.}
\label{fig:exp1}
\end{figure*}

\subsection{Experiment 2}
In Exp. 2, we investigated the effect of normal force on roughness perception of real and virtual gratings separately for 2 different spatial periods of $2.5$ and $8.5$ mm.

\subsubsection*{Stimuli}

The normal forces used in the experiment were $0.25$ N (low), $0.75$ N (medium), and $1.75$ N (high). The number of real and virtual gratings displayed separately to participants was $108$ ($3$ normal forces x $6$ repetitions x $2$ spatial periods x $3$ sessions). Hence, each session of the experiments for real and virtual gratings consisted of $36$ stimuli. 

\section{Results}

\subsection{Data Analysis}
We used the same data analysis procedure for both experiments. First, we discarded the outliers of roughness estimates in each session using Peirce's criterion. Then, we normalized roughness estimates of each participant using the method suggested in \cite{59}. We first computed the geometric mean of roughness estimates for each session, $GM_S$, and then, the geometric mean of all sessions, $GM_{TOTAL}$. Finally, we calculated the normalized estimates for each session by multiplying each estimate with $GM_{TOTAL}/GM_S$.

We used the position data acquired by IR frame to calculate the average finger speed of participants in each trial. If the actual exploration speed of a participant was $\pm30\%$ off the targeted value of $50$ mm/s, all the related data of that trial was discarded to obtain more consistent results.

\begin{figure*}[!ht]
\centering
\includegraphics[width=6.8 in]{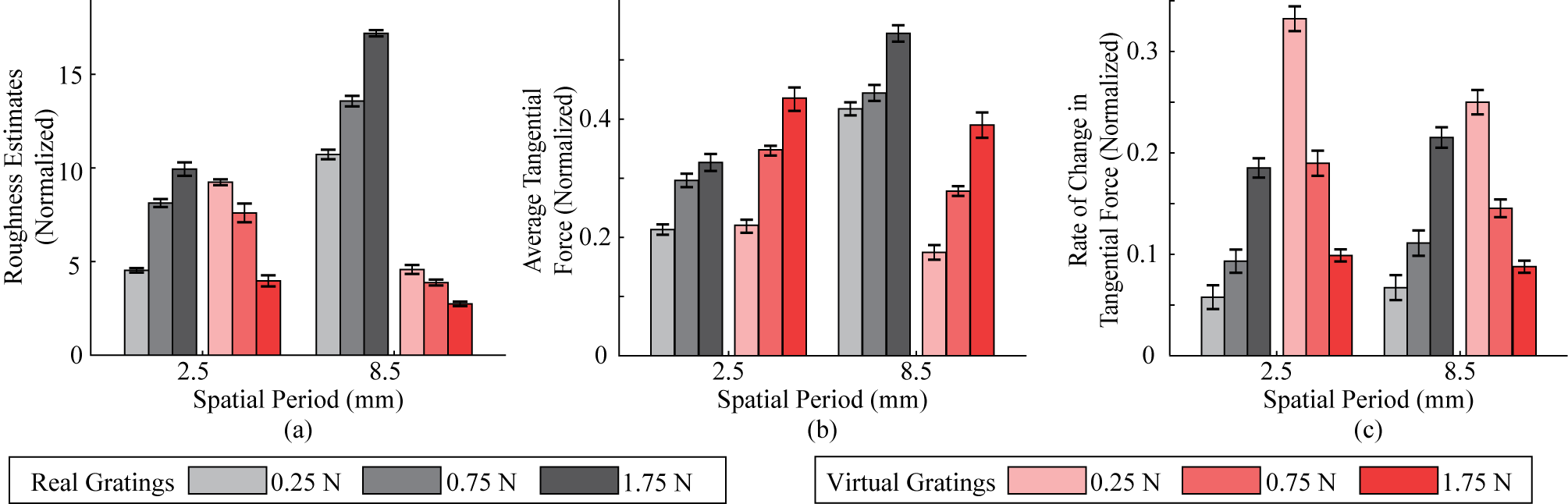}%
\caption{Results of Exp. 2; normalized roughness estimates (means and standard mean errors), average tangential force ($F_t$) and rms of rate of change in tangential force ($dF_t/dt$) for real and virtual gratings under the normal forces of $0.25$ N, $0.75$ N, $1.75$ N and targeted exploration speed of $50$ mm/s.}
\label{fig:exp2}
\end{figure*}

During each trial, both normal and tangential forces were recorded using the force sensor. A data segment of $1$ second was chosen symmetrically with respect to the location of force sensor (i.e. the mid-point of travel distance) for each trial. A bandpass filter having the cut-off frequencies of $1.25$ Hz and $1$ kHz was applied to this data segment. The following metrics were calculated for the filtered force data of each trial: average tangential force, ($F_t$), average normal force, and root mean square (rms) of rate of change in tangential force ($dF_t/dt$). Each metric was normalized between $0$ and $1$ for each session and then the average of all sessions was considered as the mean value of the participant. The average of the mean values of all participants were reported in the plots (Fig. \ref{fig:exp1} and Fig. \ref{fig:exp2}).

\subsection{Results of Experiment 1}

Table \ref{table:normalforce_speed1} shows the average exploration speeds and average normal forces applied by the participants. Normalized roughness estimates (means and standard mean errors) of real and virtual gratings are plotted as a function of spatial period for the normal force of $0.75$ N in Fig. \ref{fig:exp1}a.

\begin{table}[!h]
\caption{Average normal forces and exploration speeds in Exp. 1.}
\centering
\renewcommand{\arraystretch}{1.5}
\label{table:normalforce_speed1}
\resizebox{3.5 in}{!}{
\begin{tabular}{c|c|c|c|}
\cline{2-4}
                                                                                          & \textbf{\begin{tabular}[c]{@{}c@{}}Desired\\ Normal\\ Force\end{tabular}} & \textbf{\begin{tabular}[c]{@{}c@{}}Applied\\ Normal\\ Force\end{tabular}} & \textbf{\begin{tabular}[c]{@{}c@{}}Exploration\\ Speed\end{tabular}} \\ \hline
\multicolumn{1}{|c|}{\textbf{\begin{tabular}[c]{@{}c@{}}Real\\ Gratings\end{tabular}}}    & $0.75$ N                                                         & $0.71$ N (SD: $0.11$)                                                   & $53.05$ mm/s (SD: $7.96$)                                            \\ \hline
\multicolumn{1}{|c|}{\textbf{\begin{tabular}[c]{@{}c@{}}Virtual\\ Gratings\end{tabular}}} & $0.75$ N                                                         & $0.73$ N (SD: $0.07$)                                                   & $58.39$ mm/s (SD: $5.56$)                                            \\ \hline
\end{tabular} }
\end{table}

The results were analyzed using one-way ANOVA repeated measures. The results showed that spatial period had a significant effect on the perceived roughness of both real and virtual gratings (p $<$ $0.01$). Bonferroni corrected paired t-tests showed that the difference in roughness estimates of real gratings was significant up to the spatial periods of $4.5$ mm, as reported in the earlier studies. On the other hand, the difference in roughness estimates of virtual gratings was significant for spatial periods higher than $3.5$ mm (p $<$ $0.01$). 

The average tangential force ($F_t$) and rate of change in tangential force ($dF_t/dt$) are plotted as a function of spatial period for real and virtual gratings in Fig. \ref{fig:exp1}b and Fig. \ref{fig:exp1}c, respectively. We analyzed these results using one-way ANOVA repeated measures again. The results showed that spatial period had a significant effect on $F_t$ and $dF_t/dt$ for both real and virtual gratings (p $<$ $0.01$). Bonferroni corrected paired t-tests showed that, for real gratings, the difference in $F_t$ was significant for the spatial periods below $8.5$ mm (p $<$ $0.01$) while it was significant for the spatial periods above $3$ mm (p $<$ $0.01$) for virtual gratings. Moreover, the difference in $dF_t/dt$ was statistically significant for spatial periods below $4.5$ mm for real gratings (p $<$ $0.01$) and above $3$ mm for virtual gratings (p $<$ $0.01$).

\subsection{Results of Experiment 2}
Table \ref{table:normalforces3} shows the average exploration speeds and the average normal forces applied by participants. The normalized roughness estimates (means and standard mean errors) of real and virtual gratings were plotted as a function of normal forces ($0.25$ N, $0.75$ N, and $1.75$ N) for the spatial periods of $2.5$ mm and $8.5$ mm (see Fig. \ref{fig:exp2}a).

\begin{table}[!h]
\caption{Average normal forces and exploration speeds in Exp. 2.}
\centering
\renewcommand{\arraystretch}{1.5}
\label{table:normalforces3}
\resizebox{3.5 in}{!}{%
\begin{tabular}{c|c|c|c|}
\cline{2-4}
                                                                                                           & \textbf{\begin{tabular}[c]{@{}c@{}}Desired\\ Normal\\ Force\end{tabular}} & \textbf{\begin{tabular}[c]{@{}c@{}}Applied\\ Normal\\ Force\end{tabular}} & \textbf{\begin{tabular}[c]{@{}c@{}}Exploration\\ Speed\end{tabular}} \\ \hline
\multicolumn{1}{|c|}{\multirow{3}{*}{\textbf{\begin{tabular}[c]{@{}c@{}}Real\\ Gratings\end{tabular}}}}    & $0.25$ N                                                         & $0.26$ N (SD: $0.04$)                                                & \multirow{3}{*}{$55.67$ mm/s (SD: $6.54$)}                           \\ \cline{2-3}
\multicolumn{1}{|c|}{}                                                                                     & $0.75$ N                                                         & $0.74$ N (SD: $0.12$)                                                 &                                                                      \\ \cline{2-3}
\multicolumn{1}{|c|}{}                                                                                     & $1.75$ N                                                         & $1.66$ N (SD: $0.27$)                                                 &                                                                      \\ \hline
\multicolumn{1}{|c|}{\multirow{3}{*}{\textbf{\begin{tabular}[c]{@{}c@{}}Virtual\\ Gratings\end{tabular}}}} & $0.25$ N                                                         & $0.26$ N (SD: $0.05$)                                                 & \multirow{3}{*}{$56.70$ mm/s (SD: $7.93$)}                           \\ \cline{2-3}
\multicolumn{1}{|c|}{}                                                                                     & $0.75$ N                                                         & $0.72$ N (SD: $0.12$)                                                 &                                                                      \\ \cline{2-3}
\multicolumn{1}{|c|}{}                                                                                     & $1.75$ N                                                         & $1.60$ N (SD: $0.26$)                                                 &                                                                      \\ \hline
\end{tabular} }
\end{table}

The results were analyzed using two-way ANOVA repeated measures. The results showed that both spatial period and normal force had a significant effect on the perceived roughness of both real and virtual gratings (p $<$ $0.01$). Bonferroni corrected paired t-tests showed that, for both real and virtual gratings, the differences in roughness estimates were significant for both spatial periods under all normal forces (p $<$ $0.01$). Average tangential force ($F_t$) and rate of change in tangential force $dF_t/dt$ are plotted as a function of normal forces for both spatial periods in Fig. \ref{fig:exp2}b and \ref{fig:exp2}c, respectively. We also analyzed these results using two-way ANOVA repeated measures. The results showed that for both real and virtual gratings, spatial period and normal force, had a significant effect on $F_t$ and $dF_t/dt$. Bonferroni corrected paired t-tests showed that, for both real and virtual gratings, the differences in $F_t$ and $dF_t/dt$ were significant for all spatial periods and normal forces (p $<$ $0.01$).

\section{Discussion}

To investigate the roughness perception of real and virtual gratings, we conducted $2$ psychophysical experiments. The results showed that there are perceptual differences between real and virtual gratings.

\begin{figure}[!ht]
\centering
\includegraphics[width=3.5 in]{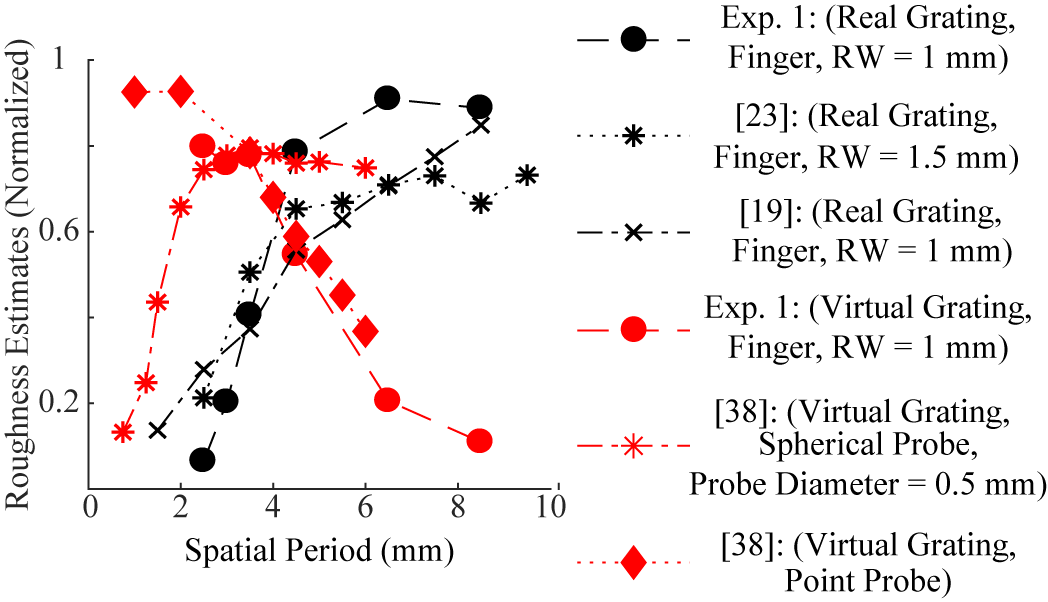}
\caption{Our results are in parallel with the literature, \cite{26,9,30}. Perceived roughness of real gratings scanned by finger and virtual gratings rendered by a force feedback device and scanned by a virtual probe having a finite size tip follow an increasing trend, while perceived roughness of virtual gratings rendered by electrovibration and scanned by finger and rendered by a force feedback device and scanned a virtual probe having an infinitely small tip follow a decreasing trend.}
\label{fig:lit}
\end{figure}

The results obtained for real gratings in the first experiment (Exp. 1) are inline with the results of earlier studies, which reported that perceived tactile roughness of macro gratings increases with increasing spatial period \cite{26,20,30,27,86}. However, this increase in perceived roughness saturates around the spatial period of $4.5$ mm \cite{26,20,30,27,9,31,39}, as observed in our study (see Fig. \ref{fig:lit}). Our results also show that the tangential force and its rate of change follow similar trends (Fig. \ref{fig:exp1}). On the other hand, the results obtained for virtual gratings in our study deviated significantly from those of the real ones. The roughness estimates of participants for virtual gratings, in contrast to the real ones, followed a decreasing trend with increasing spatial period. In fact, this is not surprising since grating height, which is important for activating SA1-afferents, cannot be rendered explicitly by electrovibration, hence, tactile perception of virtual gratings is expected to be different than that of the real ones.

In contrast to the extensive literature on real textures, the number of studies investigating the roughness perception of virtual textures is limited and mostly conducted with force feedback devices. Smith et al. \cite{94} rendered virtual gratings varying in spatial period from $1.5$ to $8.5$ mm using a force feedback device, which could only display tangential forces resisting to the planar movements, and observed a decrease in roughness perception as spatial period was increased. Unger et al. \cite{9} investigated the roughness perception of periodic virtual gratings using a force feedback device and the results suggested significant influence of virtual probe diameter. They observed a decrease in roughness perception with increasing spatial periods when virtual textures in macro size were explored with a point-probe (having an infinitely small diameter). On the other hand, there was a monotonic increase in perceived roughness for increasing spatial period from $1$ to $6$ mm for spherical virtual probes having a diameter varying from $0.25$ to $1.5$ mm in \cite{9}. 

In summary, it appears that the trend for perceived roughness of virtual gratings explored by a force feedback device, having a finite-size virtual tip, in the earlier studies resembles to that of real gratings explored by finger. On the other hand, the trend for perceived roughness of virtual gratings explored by a force feedback device, having an infinitely small virtual tip, resembles to that of the virtual gratings displayed by friction modulation explored by finger, as in our study. In electrovibration, fingerpad does not penetrate into virtual gratings since the surface topography cannot be displayed explicitly, only the tangential friction force between finger and surface is modulated periodically. As spatial period increases, the effective tangential force acting on the finger is reduced, leading to a decrease in perceived roughness.

A similar argument applies to a point probe exploring virtual gratings displayed by a force feedback device. Since surface topography can be displayed by a force feedback device and a point probe can fully penetrate into virtual grooves, the effective grating height to be overcome by the probe does not change, but the magnitude of the effective tangential force acting on the participant is reduced with increasing spatial period. As the probe size increases, the effective grating height to overcome is reduced since the probe can no longer penetrate into the gratings completely, leading to a reduction in the magnitude of effective tangential force acting on the finger. On the other hand, if the probe size is kept constant and the spatial period is increased, the probe penetrates more into the virtual grooves and the effective grating height increases until it reaches to a saturation value, as observed in tactile exploration of real gratings with a finger (Fig. \ref{fig:lit}).

In the second experiment (Exp. 2), we observed that higher normal force resulted in an increase in perceived roughness of real gratings. This result is also consistent with the results of earlier studies (\cite{20,23,28,32}) and can be also explained by our hypothesis on fingerpad penetration discussed above. As the normal force applied by the finger increases, the amount of penetration into the grooves, and hence, the effective height of the gratings to be traversed by the finger increases. As a result, the perceived roughness also increases (Table \ref{fig:discuss_figure}).

\begin{table}[!ht]
\caption{Summary of the results for Exp. 1 and Exp. 2; the up ($\uparrow$) and down ($\downarrow$) arrows in the table indicate increase and decrease in the dependent variables, respectively.}
\centering
\renewcommand{\arraystretch}{1.5}
\label{fig:discuss_figure}
\resizebox{3.5 in}{!}{%
\begin{tabular}{c}
\includegraphics[width=3.5 in]{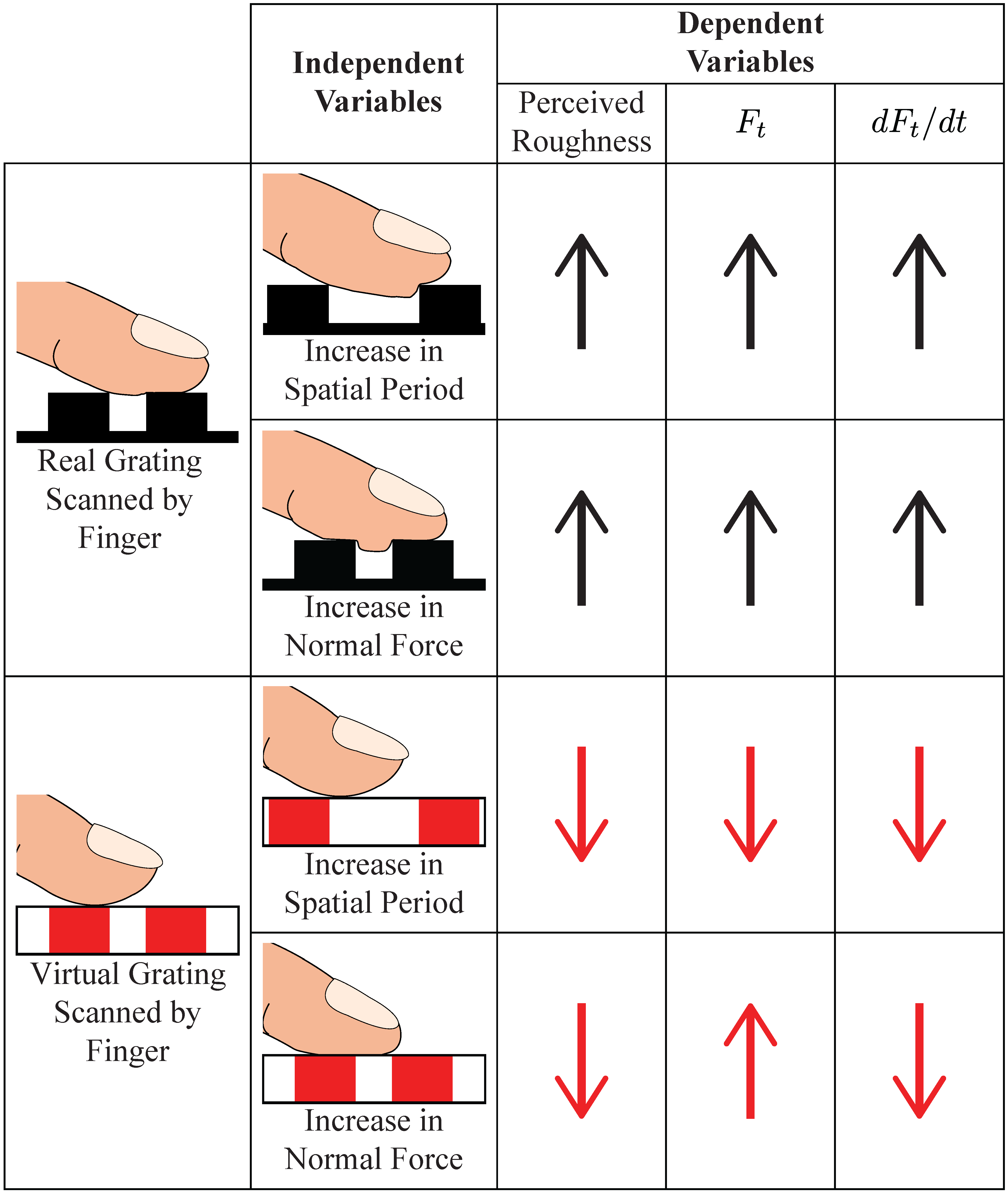}
\end{tabular} }
\end{table}

In contrast to the real gratings, higher normal force caused a decline in perceived roughness of virtual gratings in our study. Although increasing normal force increases the apparent contact area of finger, the tactile effect of electrovibration appears to decrease. It was also interesting to observe that while the tangential force acting on the participants' finger, ($F_t$), increased with higher normal force (Fig. \ref{fig:exp2}b), its rate of change, $dF_t/dt$, also increased for real gratings but decreased for virtual gratings displayed by electrovibration (Fig. \ref{fig:exp2}b). This result suggests that $dF_t/dt$ could be a better indicator of the perceived roughness for both real and virtual gratings (see Table \ref{fig:discuss_figure}). A similar conclusion was also achieved by Smith et. al. \cite{26}. They conducted psychophysical experiments with real gratings and reported that roughness perception is better correlated with the rate of change of tangential force rather than its magnitude.

\section{Conclusion}

In this study, we investigated the tactile roughness perception of real gratings made of plexiglass and virtual gratings displayed by electrovibration through a touch screen. We conducted $2$ psychophysical experiments to investigate the effect of spatial period and the normal force applied by the finger on roughness perception of real and virtual gratings. Earlier studies on real macro textures repeatedly showed that, increasing spatial period and normal force result in an increase in perceived roughness. Our results on real gratings were also inline with the earlier literature. On the other hand, the results on virtual gratings displayed by electrovibration showed that tactile roughness perception followed a decreasing trend as a function of spatial period and applied normal force. 

We argue that the difference in roughness perception of real and virtual gratings can be explained by the amount of fingerpad penetration into the gratings. This finding was consistent with the tangential force profiles recorded for both real and virtual gratings. In particular, the rate of change in tangential force ($dF_t/dt$) as a function of spatial period and normal force followed trends similar to those obtained for the perceived roughness of real and virtual gratings (Table \ref{fig:discuss_figure}). We suggest that larger spatial period and higher normal force resulted in more penetration of fingerpad into the grooves of real gratings, which in turn, resulted in an increase in the tangential force applied by the participant to overcome a ridge. Hence, the recorded tangential force, $F_t$, and its rate of change, $dF_t/dt$, increased as the spatial period was increased in real gratings. For virtual gratings, on the other hand, in which there was only friction modulation and no fingerpad penetration, the participants' finger traversed fewer number of high friction zones for larger spatial periods, and hence $F_t$ and $dF_t/dt$ decreased as the spatial period was increased.

Nonetheless, we need to clarify that, although real textures carry both spatial and temporal information, their virtual counterparts in this study were rendered based on the temporal frequency information. Therefore, the variation in exploration speed might have a stronger influence on the roughness perception of virtual textures than the real ones in our study.

In our study, we did not explicitly investigate the temporal effects of changing finger velocity on roughness perception of virtual gratings, which we plan to do so in the near future. We will also expand our study to investigate the individual effects of wavelength and duty cycle on our roughness perception of micro textures.

\section*{Acknowledgments}
The authors would like to express their gratitude to the members of RML and participants who showed great patience throughout the experiments. Moreover, A.\.{I}. would like to acknowledge Merve Adl{\i} for her unconditional support.

\bibliographystyle{IEEEtran}
\bibliography{IEEEabrv,draft}
% \newpage

\begin{IEEEbiography}[{\includegraphics[width=1in,height=1.25in,clip,keepaspectratio]{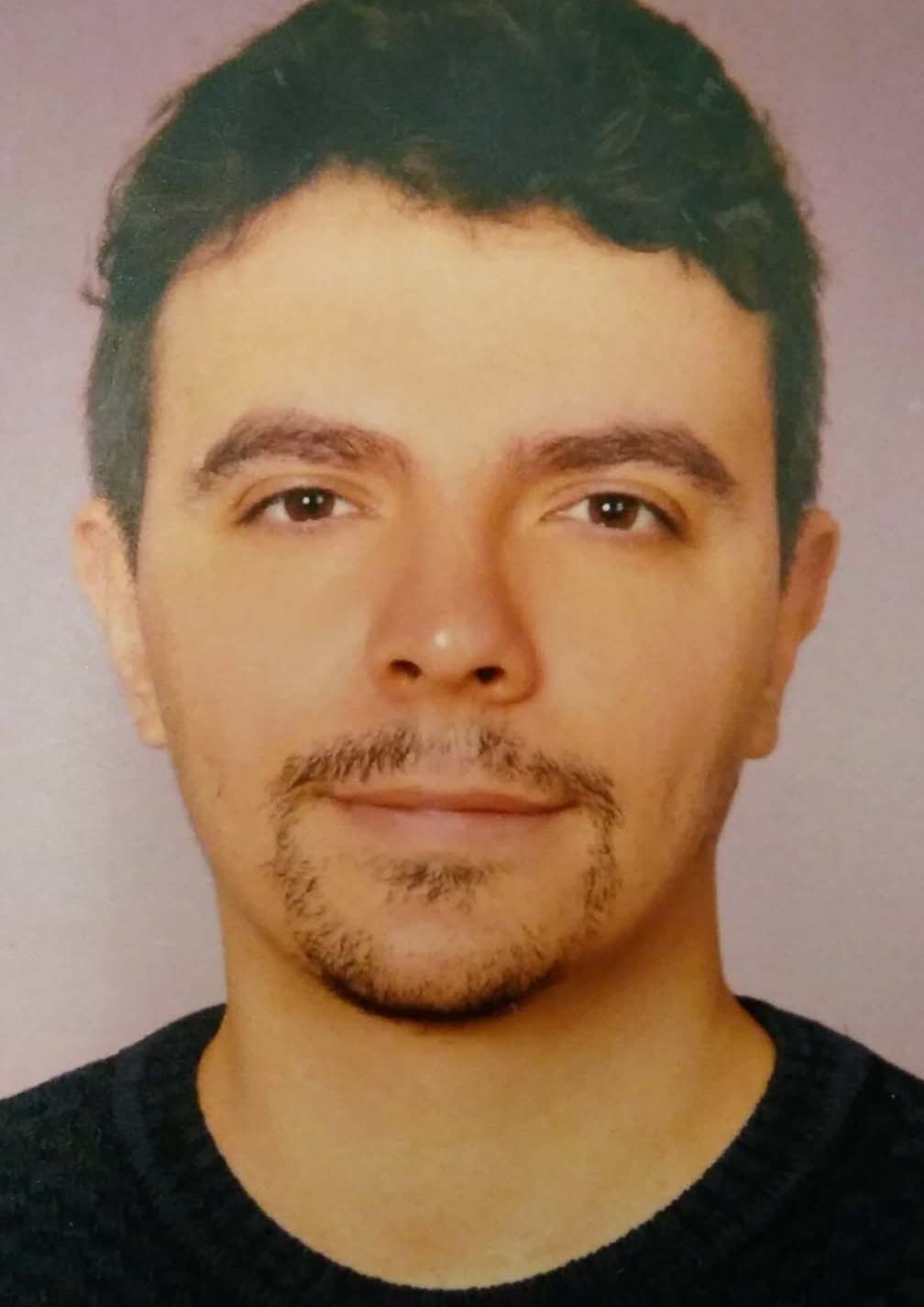} }]{Aykut \.{I}\c{s}leyen}
received the BSc degree in mechanical engineering and physics from Bogazici University, Istanbul, in 2016. He is currently a MSc student in mechanical engineering department of Koc University. He is also a member of Robotics and Mechatronics Laboratory, Koc University. His research interests include haptic interfaces, tactile perception, and psychophysics.
\end{IEEEbiography}

 %\newpage

\begin{IEEEbiography}[{\includegraphics[width=1in,height=1.25in,clip,keepaspectratio]{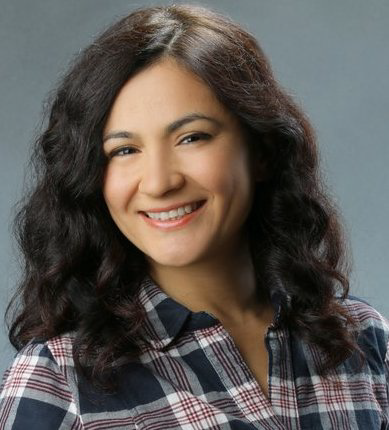}}]{Yasemin Vardar}
received the BSc degree in mechatronics engineering from Sabanci University, Istanbul, in 2010, the MSc degree in systems and control from the Eindhoven University of Technology, in 2012, and the PhD degree in mechanical engineering from Koc University, Istanbul, in 2018. She is currently a post-doctoral researcher with the Max Planck Institute for Intelligent Systems. Before starting her PhD study, she conducted research on control of high precision systems in ASML, Philips, and TNO Eindhoven. Her research interests include haptics science and applications. She is a member of the IEEE.
\end{IEEEbiography}

% insert where needed to balance the two columns on the last page with
% biographies
%\newpage

\begin{IEEEbiography}[{\includegraphics[width=1in,height=1.25in,clip,keepaspectratio]{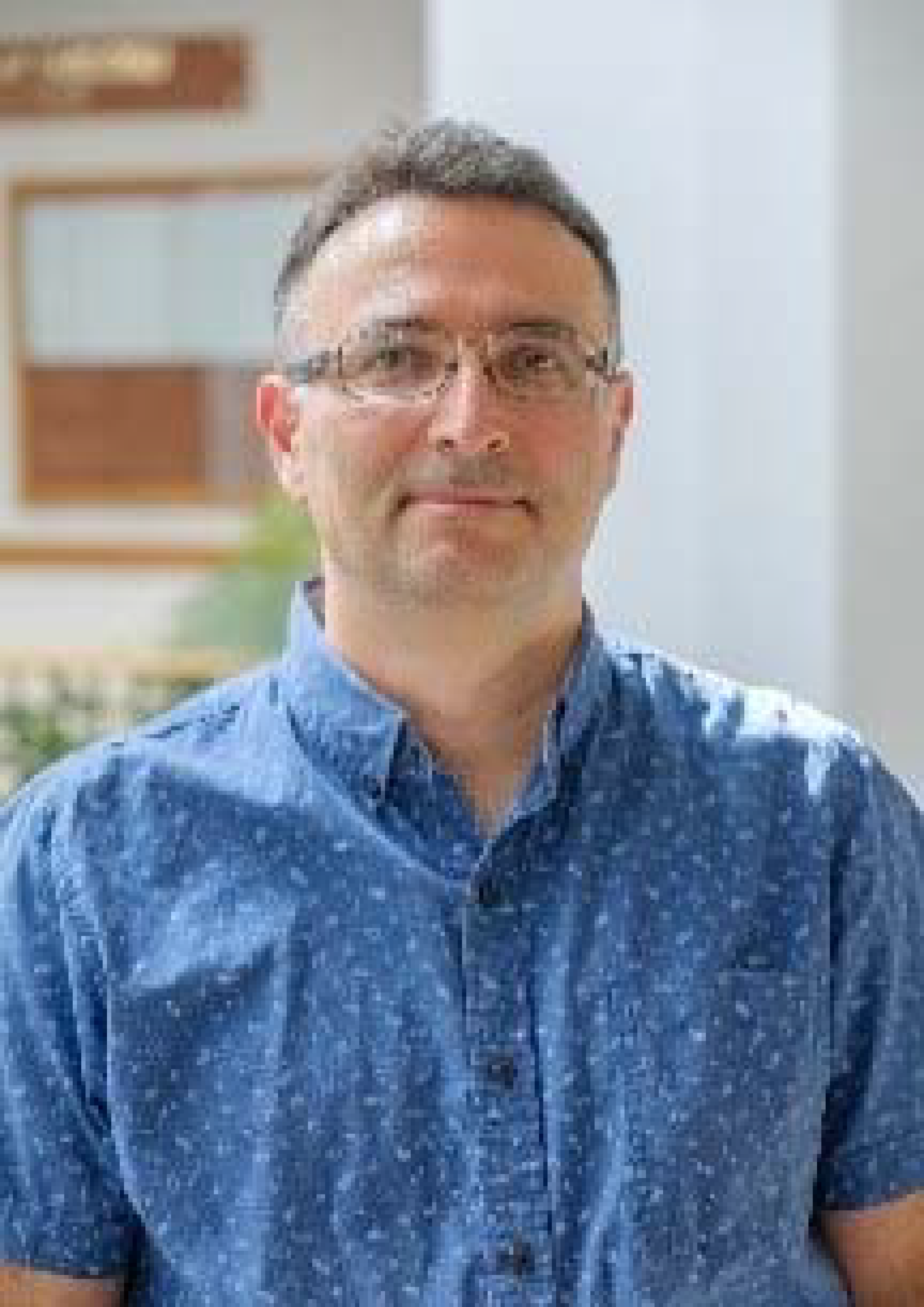}}]{Cagatay Basdogan}
received the Ph.D. degree in mechanical engineering from Southern Methodist University in 1994. He is a faculty member in the mechanical engineering and computational sciences and engineering programs of Koc University, Istanbul, Turkey. He is also the director of the Robotics and Mechatronics Laboratory at Koc University. Before joining Koc University, he worked at NASAJPL/Caltech, MIT, and Northwestern University Research Park. His research interests include haptic interfaces, robotics, mechatronics, biomechanics, medical simulation, computer graphics, and multi-modal virtual environments. He is currently the associate editor in chief of IEEE Transactions on Haptics and serves in the editorial boards of IEEE Transactions on Mechatronics, Presence: Teleoperators and Virtual Environments, and Computer Animation and Virtual Worlds journals. He also chaired the IEEE World Haptics conference in 2011.
\end{IEEEbiography}
\end{document}